\documentstyle[11pt,aaspp4,amssym]{article}

\begin{document}

\received{}
\revised{}
\accepted{}

\lefthead{M. Catelan}
\righthead{RR Lyrae luminosities: field vs. clusters}

\slugcomment{ApJ Letters, in press}

\title{Is There a Difference in Luminosity between Field and
       Cluster\\RR Lyrae Variables?}

\author{M.~Catelan}

\affil{ NASA/Goddard Space Flight Center\\
        Laboratory for Astronomy and Solar Physics, Code 681\\
        Greenbelt, MD 20771\\
        e-mail: catelan@stars.gsfc.nasa.gov
      }

\begin{abstract}
Recent {\it Hipparcos} results have lent support to the idea
that RR Lyrae variables in the halo field and in globular clusters 
differ in luminosity by $\approx 0.2$~mag. In this {\it Letter}, 
we study the pulsation properties of RR Lyraes in clusters with 
distances determined {\it via} main-sequence fitting to {\it Hipparcos} 
parallaxes for field subdwarfs, 
and compare them with the properties of field variables 
also analyzed with {\it Hipparcos}. We show that 
the period--temperature distributions for field and cluster
variables are essentially indistinguishable, thus suggesting 
that there is no significant difference in luminosity between them.
\end{abstract}

\keywords{Stars: horizontal-branch --- stars: variables: other --- Galaxy: 
          globular clusters: individual (NGC~362, M5, M68, M15, M92) 
         }


\section{Introduction}
Accurate knowledge of the Population II distance scale is one of the 
most important goals in astronomy. Upon it depends, for instance, 
the determination of the ages of globular clusters (GC's), 
and thus of a firm lower limit to the age of the Universe.

RR Lyrae variables are the natural Pop.~II ``standard candle." Several 
methods have been devised to estimate their luminosities, but a consensus 
has not yet been reached. In particular, there appears to be
a ``dichotomy" between ``faint" (i.e., short distance scale and old 
ages for the GC's) and ``bright" (long distance scale and younger GC ages) 
calibrations. Walker (1992) and Catelan (1996) provide useful references 
covering the literature as of 1995. But the noted ``dichotomy"  
has become even more clear-cut recently. Ground-based investigations 
have continued to appear supporting either the ``short" or the ``long" 
scale. Examples of the former include the Baade-Wesselink (e.g., 
Clementini et al. 1995) and statistical parallaxes (Layden et al. 1996; 
Popowski \& Gould 1998) analyses of field RR Lyraes. An example of the 
latter has been provided by the extensive analysis of the 
variables in M15 by Silbermann \& Smith (1995). 
The ``persistent" nature of such a ``dichotomy" has led some authors 
(e.g., VandenBerg, Bolte, \& Stetson 1996; Sweigart 1998) to 
speculate that {\it there might exist a real difference in luminosity 
between field and cluster RR Lyrae variables}. That was based in part on the 
(somewhat uncertain) Baade-Wesselink results of Storm, Carney, \& Latham 
(1994) for a few RR Lyrae variables in the GC's M5 and M92 and field 
counterparts of comparable metallicity. 

These speculations notwithstanding, there has been widespread belief that, once
the {\it Hipparcos} satellite parallax results became available, we would 
finally be able to decide between the ``short" and ``long" RR Lyrae distance 
scales. However, that turned out {\it not} to be the case. Based upon  
{\it Hipparcos} parallaxes of field subdwarfs and main-sequence fitting to 
GC's with well-defined deep color-magnitude diagrams, Gratton 
et al. (1997) and Reid (1997, 1998) have strongly claimed that the majority 
of the GC's in their samples are substantially farther away than previously 
estimated using ground-based parallaxes (but see Pont et al. 1998). 
Similarly, McNamara (1997b) has concluded that the {\it Hipparcos}
parallaxes of field SX Phoenicis variables favor the ``long" GC distance
scale. These claims were supported by {\it Hipparcos} data for Cepheids 
(Feast \& Catchpole 1997; Madore \& Freedman 1997; see also the latest 
ground-based results by Laney 1998) and Miras (van Leeuwen et al. 1997), 
applied to the Large Magellanic Cloud (LMC). The ``long" distance 
to the LMC is supported by the latest analysis of the SN1987a ring (Panagia, 
Gilmozzi, \& Kirshner 1998; but see Gould \& Uza 1998). On the other hand, 
Gratton (1998) has analyzed {\it Hipparcos} data for field horizontal-branch 
(HB) stars including three RR Lyrae variables, and found that the faint HB 
luminosity scale was preferred. Fernley et al. (1998) and Tsujimoto, Miyamoto, 
\& Yoshii (1998) have also reported, based on {\it Hipparcos} data for
field RR Lyraes, luminosities which are consistent with the corresponding
ground-based analyses. As argued by Gratton, {\it the Hipparcos results 
thus seem to favor the existence of an intrinsic difference in luminosity 
(by $\approx 0.2$~mag) between GC and field RR Lyraes}. 

However, no independent tests have thus far been applied to verify this. 
As is well known, {\it RR Lyrae pulsation properties depend strongly on 
their luminosities}.
The purpose of this {\it Letter} is to employ such properties
to constrain the difference in luminosity 
between field and GC variables. {\it Only GC's and field stars analyzed 
with Hipparcos will be covered.} We begin in Sec.~2 by discussing the 
employed methods for deriving RR Lyrae temperatures. In Sec.~3, the 
selection criteria we have adopted are described. In Sec.~4, we 
demonstrate that GC variables do not show substantial period shifts 
with respect to field variables of similar metallicity, as opposed 
to what would be expected if there were an intrinsic luminosity difference 
between them. Finally, our results are critically discussed in 
Sec.~5.\footnote{We emphasize that the purpose of the present work 
is to perform a period-shift analysis at fixed temperature {\it and}  
metallicity. Thus, a careful analysis of the Sandage (1993) 
period-shift effect lies outside the scope of this {\it Letter}.}

\section{Estimating RR Lyrae Temperatures}
\subsection{The Carney, Storm, \& Jones (1992a) Approach}
In their Baade-Wesselink analysis of field RR Lyraes, Carney et al. (1992a,
hereafter CSJ92) compiled parameters for a number of variables, 
including temperatures derived from near-infrared colors. 
Analyzing possible correlations in their database, they concluded that 
a simple equation exists [their eq.~(16)] relating the ``equilibrium 
temperature" $T_{\rm eq}$, blue amplitudes $A_B$, pulsation periods $P$, 
and metallicities [Fe/H] 
for ab-type RR Lyraes. This relationship formed the basis for their 
discussion of the period-shift effect, and will be adopted here as a 
first means of estimating temperatures.
\subsection{The Catelan, Sweigart, \& Borissova (1998) Approach}
Catelan et al. (1998, hereafter CSB98) have recently reanalyzed
temperatures based on the CSJ92 data. They argued that a 
relationship involving only $T_{\rm eq}$, $A_B$, and [Fe/H] 
would be safer to adopt in period-shift analyses than CSJ92's 
(Sec.~2.1), since period shifts caused by luminosity variations could 
easily be misinterpreted as being due to temperature variations. The idea of 
employing $A_B$ values to determine $T_{\rm eq}$ (cf. Sandage 1981a,b)
is supported by Jones et al. (1992), who state that ``$\ldots$~at a 
fixed metallicity, it is likely that {\it relative} $A_B$ values are reliable 
indicators of {\it relative} temperatures."

%
\begin{figure}[t]
 \plotfiddle{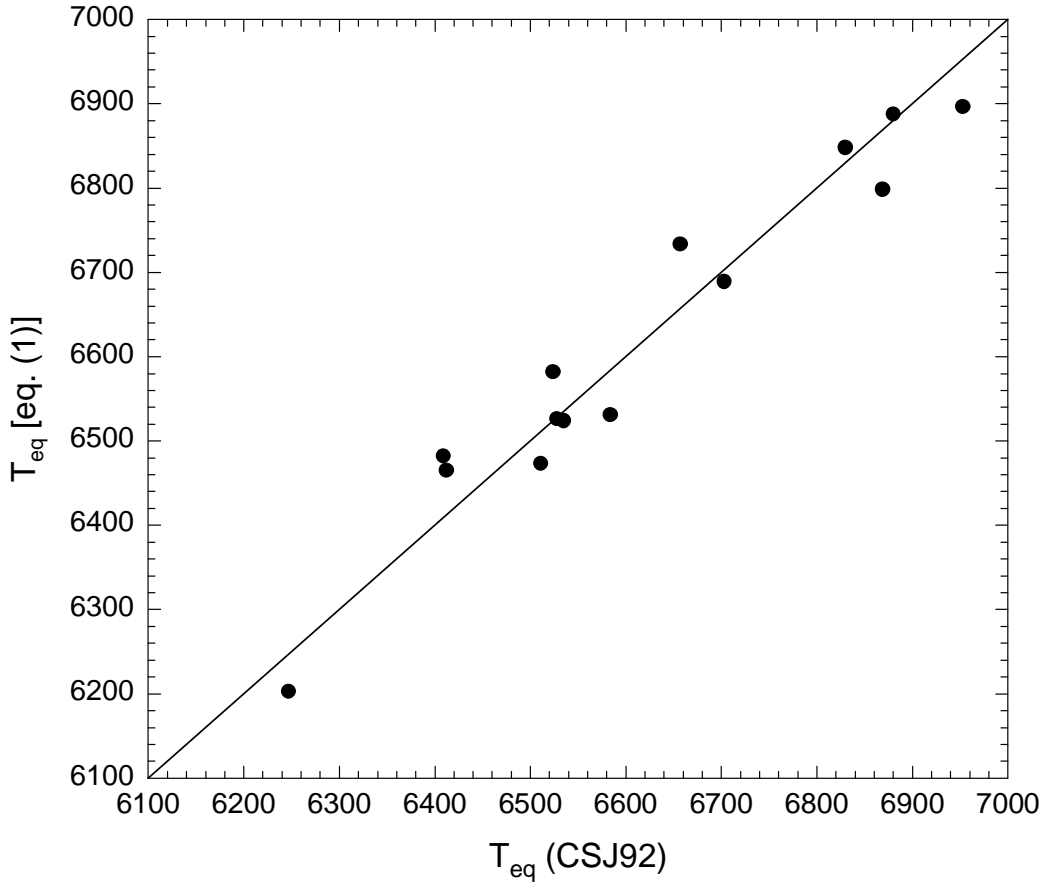}{11.5cm}{0}{90}{90}{-275}{-185}
 \caption[]{Equilibrium temperatures from eq.~(1) are plotted
   as a function of the values actually {\it derived} by CSJ92 (and 
   references therein).}
\end{figure}

We have rederived the CSB98 relationship for $T_{\rm eq}$ for the 
same selection criteria and parameters used in our  
period-shift analysis of the {\it Hipparcos} sample (Sec.~3). Thus, the star 
SW~And was also removed from the CSJ92 database, because it presents the 
Blazhko effect. Furthermore, the $A_B$ values from Blanco (1992) were 
adopted. (The differences are generally small, with the exception of 
DX~Del, for which Blanco's $A_B$ is larger by $0.28$~mag.)

Our new relationship for $\Theta_{\rm eq} = 5040/T_{\rm eq}$ thus reads:
\begin{equation}
 \Theta_{\rm eq} = (0.868\pm 0.014) - (0.084\pm 0.009)\, A_B - 
                   (0.005\pm 0.003)\, {\rm [Fe/H]},
\end{equation}
\noindent with a multiple correlation coefficient $r = 0.97$ and a rms 
deviation of $\simeq 40$~K. Fig.~1 shows that this relationship does 
provide a superb match to the CSJ92 equilibrium temperatures.

\section{RR Lyrae Stars: Adopted Samples}
In the present section, we lay out the selection criteria employed in our 
analysis.
\subsection{Field RR Lyrae Stars}
We have retrieved the list of 125 variables employed by Tsujimoto et al. 
(1998) in their {\it Hipparcos}-based analysis of field RR Lyraes, as
kindly supplied by Dr.~T. Tsujimoto.

We have selected stars from this sample according to the following criteria.
i)~{\it Reliable classification as ab-type RR Lyrae stars}: 
Variables whose RRab Lyrae nature has been questioned by Schmidt, Chab, \& 
Reiswig (1995) or Fernley \& Barnes (1997) were dismissed;
ii)~{\it Well-behaved light curves}: Stars which Blanco (1992) or 
Schmidt et al. pinpointed as Blazhko variables were discarded;
iii)~{\it Metallicity values available from Layden}: Stars for which 
Layden et al. (1996) do not provide metallicity values were not considered. 
We have also removed from the list all variables for which metallicities 
are based on Hemenway's (1975) measurements, since we consider the 
corresponding Layden et al. [Fe/H] values quite uncertain;
iv)~{\it Blue amplitudes and pulsation periods available from Blanco (1992)}.
\subsection{Cluster RR Lyrae Stars}
Among the 12 GC's which have had distances determined using 
{\it Hipparcos} parallaxes for field subdwarfs, only 5 contain a sufficiently 
large number of RR Lyrae variables to justify their inclusion in the present 
analysis: NGC~362, M5, M68, M15, and M92. Since the 
Layden et al. (1996) field RR Lyrae metallicities are tied in to the 
Zinn \& West (1984) abundance scale, we decided to adopt the [Fe/H] 
entries of Harris' (1996) catalogue for consistency. Although the Zinn \& West 
scale has been seriously questioned by Gratton et al. (1997 and references 
therein), this is of minor relevance for the present purposes, since our goal 
is to perform a period-shift analysis at {\it fixed metallicity}. Likewise, 
the criticism of McNamara (1997a) of the near-infrared temperatures 
is of secondary relevance for us.

%
\begin{figure}[t]
 \plotfiddle{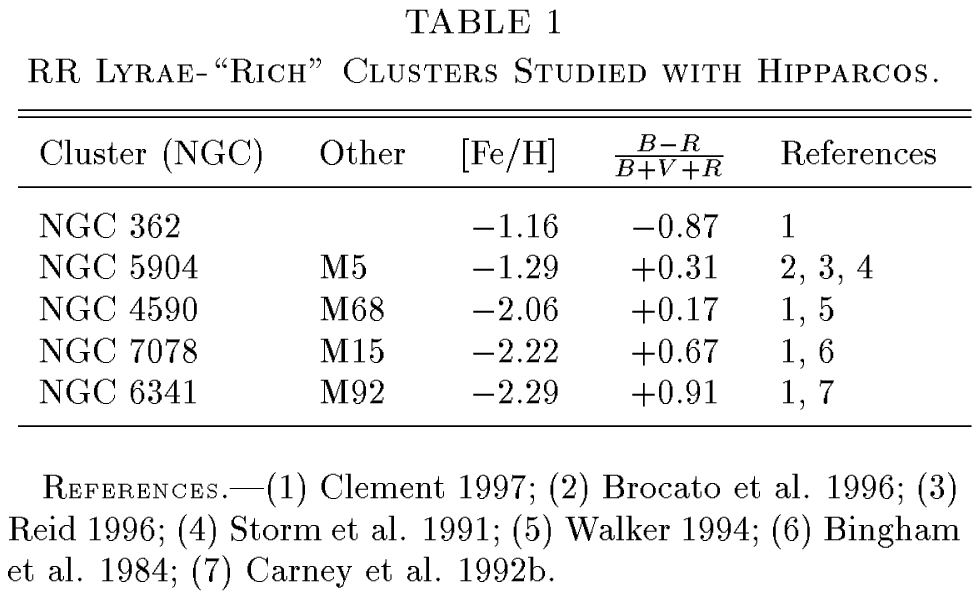}{5cm}{0}{80}{80}{-245}{-240}
\end{figure}

The adopted sources of information for the GC RR Lyrae variables are provided 
in Table~1, along with the cluster [Fe/H] and Lee-Zinn HB morphology parameter 
(both from Harris 1996). According to such HB types, the only GC 
for which evolution away from the blue zero-age HB may bias the period-shift 
analysis is M92. As with the field star sample, Blazhko variables were 
discarded---as were those suspected to be non-cluster members.

\section{Period-Shift Analysis: Clusters versus Field}
Table~1 shows that our GC sample divides into two  
metallicity bins, with ${\rm [Fe/H]} \approx -1.2$ and $\approx -2.2$. 
We have thus split the comparison 
between GC and field variables into two metallicity regimes. For the 
more metal-rich end, we employ all field RR Lyraes (25 stars) falling 
in the range $-1.50 \leq {\rm [Fe/H]} \leq -0.95$ which have 
passed our selection criteria (Sec.~3.1); at the metal-poor end, we restrict 
the sample to the variables with ${\rm [Fe/H]} \leq -1.85$ 
(10 stars). The resulting $\log\,P - \log\,T_{\rm eq}$ diagrams using 
temperatures derived as in Sec.~2.1 and Sec.~2.2 are shown in Figs.~2 and 
3, respectively.

%
\begin{figure}[b]
 \epsscale{1.20}
 \plotfiddle{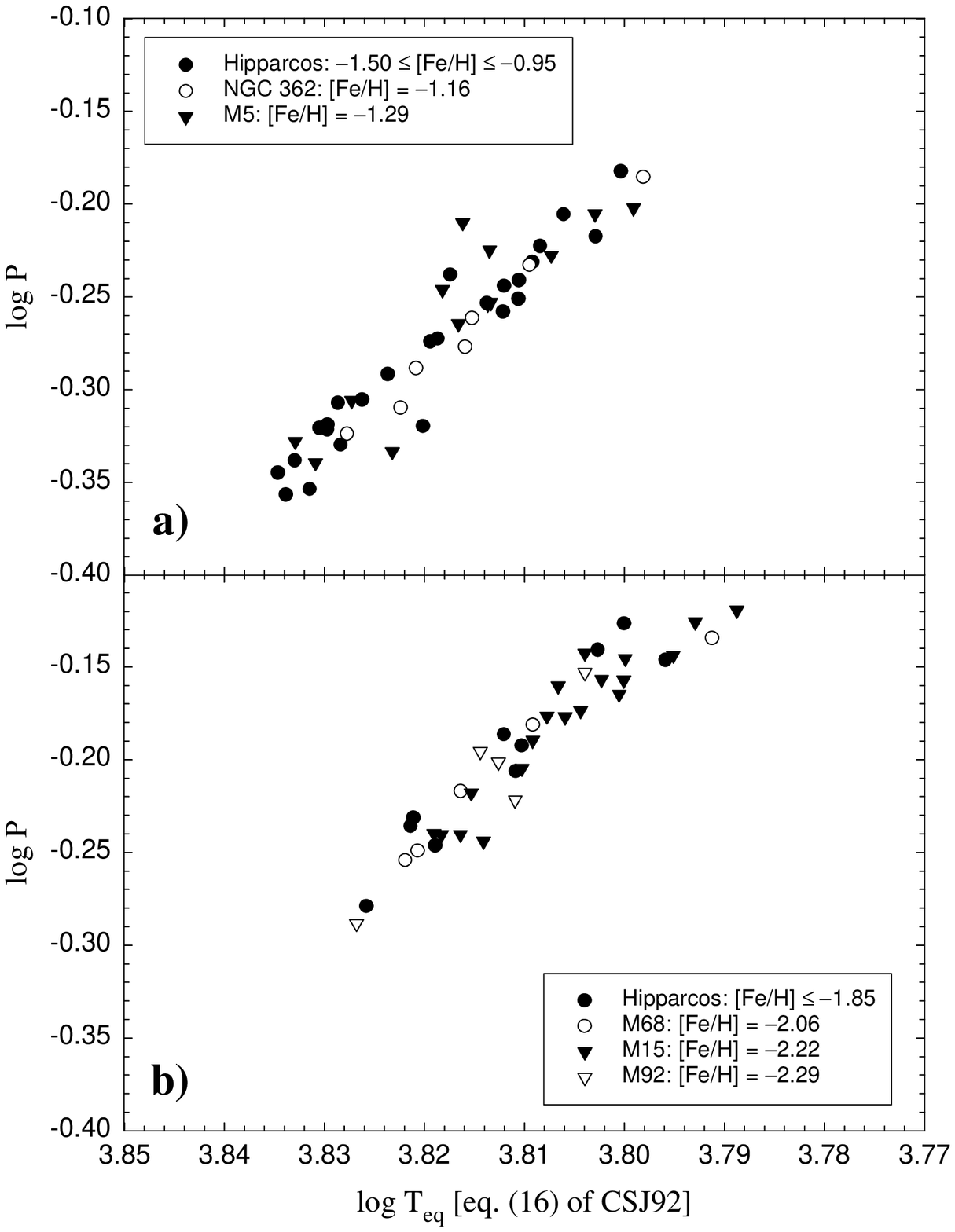}{17cm}{0}{90}{90}{-280}{-100}
 \figurenum{2}
 \caption[]{a) Period-temperature diagram for RR Lyraes 
   in the GC's NGC~362 ($\circ$) and M5 ($\blacktriangledown$),
   compared with variables investigated with {\it Hipparcos} ($\bullet$) 
   in the indicated metallicity range. b) As in panel (a), except that 
   the more metal-poor regime is studied. Variables in M68 ($\circ$), 
   M15 ($\blacktriangledown$) and M92 ($\triangledown$) are shown. 
   Temperatures have been derived from eq.~(16) of CSJ92.}
\end{figure}

%
\begin{figure}[b]
 \epsscale{1.20}
 \plotfiddle{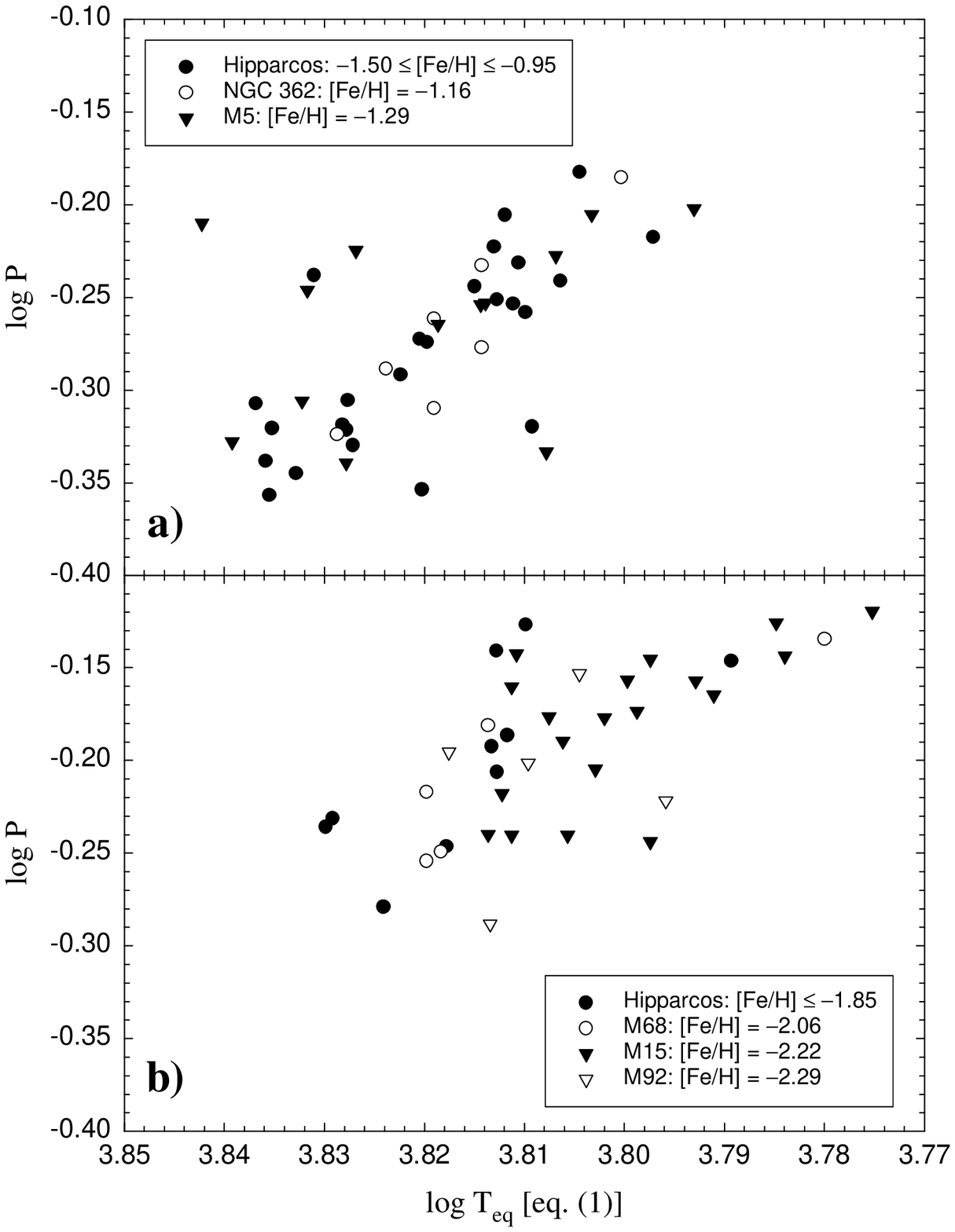}{17cm}{0}{90}{90}{-280}{-100}
 \figurenum{3}
 \caption[]{As in Fig.~2, except that temperatures have been
   derived from our eq.~(1).}
\end{figure}

The scatter is substantially larger in Fig.~3
than in Fig.~2. This, however, should not be taken as evidence that
eq.~(1) is less satisfactory at estimating $T_{\rm eq}$ values than
eq.~(16) of CSJ92. As previously argued (Sec.~2.2), CSJ92's 
relationship, by including a {\it period} term,
can ``mask" luminosity variations at a {\it fixed}
$T_{\rm eq}$, misinterpreting them as temperature variations. 
Thus, eq.~(16) of CSJ92 artificially
{\it drives} a tight $P-T_{\rm eq}$ distribution for a sample of stars
with intrinsic luminosity scatter. Our eq.~(1) does
not have this bias, being more suitable for detecting 
luminosity variations at a given $T_{\rm eq}$. Consider, for instance,
V9 in 47~Tuc ($P = 0.737$~d), which is brighter than field RR Lyraes
of similar metallicity by $\approx 0.6$~mag (cf. Fig.~9 in Storm et 
al. 1994). Eq.~(16) of CSJ92 underestimates V9's 
$T_{\rm eq}$ by $\simeq 600$~K, while the underestimate from eq.~(1) 
is only $\simeq 180$~K. In fact, Marconi's (1997, priv. comm.) pulsation 
models show temperatures to be quite insensitive to $L$ at 
fixed {\it blue} amplitude over a range in $M_{\rm bol}$ of 0.75~mag 
and for $0.2 \lesssim A_B \lesssim 2.0$. 
In addition, SS~Leo, which has 
ill-determined physical parameters, was not discarded
by CSJ92 when deriving their eq.~(16). Excluding this star from 
the CSJ92 sample, we find a reduction in the $\log\,P$ coefficient of 
their eq.~(16) by $\approx$~a factor of two, and an increase (in absolute 
value) in the corresponding $A_B$ coefficient by a similar factor. The 
$T_{\rm eq}$ value adopted by CSJ92 for this star, $\sim 6400$~K, differs 
from the one expected on the basis of eq.~(1) by $\simeq 300$~K---a factor 
of $\approx 3$ larger than the one for the largest-deviating star in our Fig.~1.

Figs.~2 and 3 show that {\it there 
is no detectable difference in period-shift properties between the studied 
field and cluster RR Lyraes, either at the metal-poor or at the more 
metal-rich end}---irrespective of the approach used to estimate $T_{\rm eq}$. 
If Gratton's (1998) suggestion were correct and the GC variables 
were brighter by $\approx 0.2$~mag, we would expect to see a difference 
as large as $\Delta\log\,P \approx +0.067$ at fixed $T_{\rm eq}$ between GC 
and field stars (Catelan 1996), which is most decidedly {\it not} present 
in our diagrams.

Other interesting conclusions that may be drawn from 
Figs.~2 and 3, but which we shall not discuss in the present 
{\it Letter}, are: i)~There seems to be no offset in the $P - T_{\rm eq}$ 
diagrams between GC's with widely different HB types but similar [Fe/H] 
(cf. Catelan 1994); ii)~Metal-poor RRab Lyraes may have a cooler 
$T_{\rm eq}$ cutoff than the metal-rich ones (Sandage 1993); iii)~There 
may be an offset of $\Delta\log\,P \approx +0.05$ at constant $T_{\rm eq}$ 
($\Rightarrow$ $\Delta M_{\rm bol}$ $\approx 0.15$ mag for fixed mass) 
between the metal-poor and the more metal-rich RR Lyraes.

\section{Discussion}
{\it The present analysis does not substantiate Gratton's (1998) suggestion, 
based on Hipparcos results, that there is a difference in 
luminosity between GC and field RR Lyrae variables}, showing instead  
that they have essentially the same distribution
in the $P - T_{\rm eq}$ plane, both at the metal-poor and at the more 
metal-rich ends.

Does this imply that there is really {\it no} difference between GC and field 
HB stars? Not necessarily. In fact, at ${\rm [Fe/H]} > -1$, Sweigart \& 
Catelan (1998) found (following the same approach as in the present 
{\it Letter}) substantial differences between field and (some) GC RR Lyraes 
(see also Storm et al. 1994 and Layden 1995). Moreover, it should be noted
that: i)~The $A_B - T_{\rm eq}$ diagram (and possibly even 
spectroscopically-derived metallicities) 
may be sensitive to the helium abundance $Y$, so that an additional, 
$Y$-dependent term may be needed to put eq.~(1) on a firmer basis (CSB98). 
In any case, available models suggest that it would not be possible 
for differences in $Y$ between field and GC stars to be consistent with
both a luminosity difference of $\approx 0.2$~mag and the remarkable
overlapping in the $P - T_{\rm eq}$ plane found in Figs.~2 and 3; 
ii)~As well known, field red giants do not seem to show signatures 
of non-canonical deep mixing, whereas some GC's do (cf. Kraft et al. 1997). 
It might be worth examining whether RR Lyrae temperatures and amplitudes  
might be sensitive to their (inherited) abundance anomalies; 
iii)~As pointed out by Sweigart (1998), the stars which are more likely to 
be affected by ``helium mixing" during the red giant branch phase are the 
blue-HB and extreme-HB stars, not the cooler RR Lyraes.

The present results (see also Fernley 1993) provide motivation for 
searching for systematic errors in methods employed to estimate the 
distances of GC's (esp. main-sequence fitting) and the luminosities of 
RR Lyrae stars (esp. the statistical parallaxes and Baade-Wesselink 
methods). Unless some ``cosmic conspiracy" is leading to the remarkable 
agreement between field and cluster stars in Figs.~2 and 3, 
{\it the ``long" and ``short" Pop.~II distance scales cannot be reconciled 
in the way suggested by Gratton (1998)}. We cannot tell whether the ``long" 
or the ``short" scale should be preferred on the basis of a comparison with 
evolutionary models, due to extant systematic uncertainties in the 
empirical RR Lyrae temperatures (McNamara 1997a). However, the LMC provides 
a means of estimating RR Lyrae luminosities (Walker 1992), and several methods 
seem to favor the ``long" distance scale (implying brighter RR Lyraes and 
younger GC ages) over the ``short" one.

\acknowledgments
The author would like to thank C. Cacciari, A. Sweigart, D. VandenBerg, 
W. Landsman and the referee for useful suggestions, and M. Marconi and 
T. Tsujimoto for providing relevant information. This work was performed 
while the author held a National Research Council--NASA/GSFC Research 
Associateship.

\end{document}